# Enhancing Trust Through Standards: A Comparative Risk-Impact Framework for Aligning ISO AI Standards with Global Ethical and Regulatory Contexts

Sridharan Sankaran
Research and Innovation Group
Tata Consultancy Services
Chennai, India
0009-0005-1341-6384

*Abstract*—As artificial intelligence (AI) reshapes industries and societies, ensuring its trustworthiness—through mitigating ethical risks like bias, opacity, and accountability deficits—remains a global challenge. International Organization for Standardization (ISO) AI standards, such as ISO/IEC 24027 and 24368, aim to foster responsible development by embedding fairness, transparency, and risk management into AI systems. However, their effectiveness varies across diverse regulatory landscapes, from the EU's risk-based AI Act to China's stability-focused measures and the U.S.'s fragmented state-led initiatives. This paper introduces a novel Comparative Risk-Impact Assessment Framework to evaluate how well ISO standards address ethical risks within these contexts, proposing enhancements to strengthen their global applicability. By mapping ISO standards to the EU AI Act and surveying regulatory frameworks in ten regions—including the UK, Canada, India, Japan, Singapore, South Korea, and Brazil—we establish a baseline for ethical alignment. The framework, applied to case studies in the EU, US-Colorado, and China, reveals gaps: voluntary ISO standards falter in enforcement (e.g., Colorado) and undervalue region-specific risks like privacy (China). We recommend mandatory risk audits, region-specific annexes, and a privacy-focused module to enhance ISO's adaptability. This approach not only synthesizes global trends but also offers a replicable tool for aligning standardization with ethical imperatives, fostering interoperability and trust in AI worldwide. Policymakers and standards bodies can leverage these insights to evolve AI governance, ensuring it meets diverse societal needs as the technology advances.

*Keywords*— AI Standards ,Trustworthiness, Regulatory frameworks, Risk impact, AI Ethics

## I. INTRODUCTION

### A. Background and Context

Artificial intelligence (AI) has emerged as a transformative force, reshaping industries such as healthcare, finance, manufacturing, and education through capabilities like autonomous decision-making and predictive analytics. Its integration into critical societal functions offers immense potential for innovation and productivity, yet it also introduces significant ethical, legal, and societal challenges. Issues such as algorithmic bias, lack of transparency, and accountability deficits have been widely documented, particularly in high-stakes domains like law enforcement and healthcare [1]. For instance, biased AI systems in hiring have disproportionately excluded marginalized groups, raising concerns about fairness and equity [2]. As AI's societal footprint grows, ensuring its trustworthiness—defined as reliability, safety, and ethical alignment—becomes paramount [3]. Without robust governance, the risks of harm, including privacy violations and unintended consequences, threaten public trust and adoption [4]. These challenges underscore the urgency of developing frameworks that balance technological advancement with societal well-being, a task increasingly addressed through standardization and regulation.

### B. The Role of Standards in AI

Standards play a pivotal role in mitigating AI's risks by providing a shared framework for consistency, interoperability, and ethical accountability across industries [5]. They serve as a bridge between governance and technical innovation [6] on one side and societal expectations on the other, embedding principles like transparency, fairness, and safety into AI systems [7]. The International Organization for Standardization (ISO), through its ISO/IEC JTC 1/SC 42 committee [8], has developed guidelines addressing data quality, bias, and risk management, aiming to foster trustworthy AI. For example, ISO/IEC 24027 [9] offers tools to assess and reduce unwanted bias, directly tackling ethical concerns. Beyond technical specifications, standards signal a commitment to ethical values, enabling developers to align AI with human rights and regulatory requirements. However, their effectiveness depends on adoption and adaptability to diverse global contexts, where regulatory priorities—such as the EU's risk-based approach or China's focus on social stability—differ significantly [10]. This variability highlights the need for a nuanced evaluation of standards' impact on ethical risks.

### C. Need for a New Analytical Approach

While ISO standards provide a universal foundation for responsible AI, their application across heterogeneous regulatory regimes reveals limitations. Existing mappings, such as those aligning ISO standards with the EU AI Act, focus narrowly on compliance within specific frameworks, overlooking gaps in addressing region-specific ethical risks [11]. For instance, the voluntary nature of ISO/IEC 23894 on risk management [12] may suffice in the EU's structured environment but falters in the U.S., where state-level mandates like Colorado's AI Act [13] contrast with federal deregulation [14]. Similarly, China's emphasis on social harmony requires privacy safeguards beyond ISO's current scope. These disparities suggest that a static, one-size-fits-all approach to standardization is insufficient. This paper proposes a Comparative Risk-Impact Assessment Framework to address this gap, offering a systematic method to evaluate and enhance ISO standards' effectiveness in mitigating ethical risks across diverse global contexts, thereby advancing their role in building trustworthy AI.

### D. Scope and Objectives of the Paper

This paper aims to enhance trust in AI systems by synthesizing global regulatory trends and ISO standardization efforts while introducing an original contribution: the Comparative Risk-Impact Assessment Framework. Beyond mapping ISO standards to the EU AI Act, we develop and apply this framework to assess how well these standards address ethical risks—such as bias, transparency, and accountability—in varying regulatory landscapes, including the EU, U.S. states, China, and other key regions. By identifying gaps and proposing targeted enhancements to ISO standards, we seek to strengthen their adaptability and impact, fostering responsible AI development worldwide. The scope encompasses an overview of ISO's role, a survey of emerging AI regulations, a detailed standards mapping, and the framework's application to case studies, culminating in actionable recommendations. This approach bridges theoretical standardization with practical governance, offering a novel tool for aligning AI innovation with ethical and societal imperatives.

## II. THE IMPORTANCE OF AI STANDARDIZATION

### A. Promoting Interoperability and Ethical Integration

AI standardization fosters interoperability by establishing a common technical and ethical language, enabling seamless collaboration across systems and stakeholders [5]. This consistency is critical as AI applications—like autonomous vehicles or medical diagnostics—span borders and industries, requiring compatible frameworks to ensure functionality and safety [7] Beyond technical integration, standards embed ethical principles such as fairness and inclusivity into AI design, reducing risks of harm to vulnerable populations [15]. For instance, interoperable standards can prevent fragmented AI ecosystems that exacerbate inequities, as seen in biased facial recognition systems disproportionately misidentifying minorities [16]. By harmonizing technical and ethical dimensions, standardization lays the groundwork for innovation that aligns with societal values, a foundation this paper builds upon to assess global regulatory alignment.

### B. Enhancing Trust through Ethical Safeguards

Standards enhance trust by providing ethical safeguards that mitigate AI's inherent risks, such as bias, discrimination, and opacity [17]. Defined protocols—like ISO/IEC 24027 [9] for bias assessment—offer measurable criteria to ensure reliability and fairness, addressing public concerns about "black box" AI systems [18]. Transparency requirements, a cornerstone of many standards, compel developers to disclose decision-making processes, building confidence among users and regulators [4]. Research shows that trust in AI increases when ethical safeguards are visible and enforceable, as evidenced by public backlash to unregulated AI in hiring [2]. These safeguards are vital for widespread adoption, yet their effectiveness varies across regulatory contexts, necessitating a comparative approach to evaluate their impact.

### C. Supporting Ethical and Responsible AI Development

Standardization supports ethical and responsible AI development by embedding societal values into technical processes, addressing concerns like privacy and accountability [19]. Standards mandate practices that reduce ethical risks—for example, ISO/IEC 5259 [20] ensures data quality to prevent biased outcomes in machine learning. This alignment with human rights and fairness is critical in sensitive applications, such as healthcare AI, where errors can perpetuate harm [21]. By providing a structured framework, standards guide developers toward accountability, ensuring AI serves the public good rather than amplifying existing inequalities [22]. However, the diversity of global ethical priorities highlights the need for adaptive standards, a challenge this paper tackles through its proposed framework.

### D. Facilitating Global Harmonization

AI standardization facilitates global harmonization by creating consistent guidelines that transcend national boundaries, fostering equitable access and collaboration [5]. International standards, such as those from ISO/IEC JTC 1/SC 42, reduce trade barriers and align AI practices with universal ethical norms, leveling the playing field for innovation [10]. This is particularly relevant as countries like the EU and China pursue divergent regulatory paths—risk-based versus control-focused—yet share common goals of fairness and safety [14]. Harmonization ensures that ethical disparities, such as unequal privacy protections, are minimized, benefiting global society. Yet, achieving true harmonization requires standards to adapt to regional nuances, a gap this paper's framework seeks to address.

### E. Guiding Ethical Innovation and Regulatory Compliance

Standards guide ethical innovation by identifying benchmarks for improvement, such as fairness and inclusivity, while ensuring compliance with evolving regulations [7]. They provide evaluation criteria—like ISO/IEC 25059's [23] quality model—that help organizations meet legal and ethical mandates, reducing risks of misuse. As regulatory frameworks, such as the EU AI Act, impose stringent requirements on high-risk AI, standards serve as a compliance roadmap, bridging innovation and accountability [11]. However, their voluntary nature can limit enforceability in less regulated contexts, like the U.S., where state laws outpace federal action [13]. This tension underscores the need for a comparative risk-impact analysis to enhance standards' global applicability, a core contribution of this work.

## III. OVERVIEW OF ISO AND ITS ROLE IN AI STANDARDIZATION

### A. Introduction to ISO

The International Organization for Standardization (ISO), founded in 1947, is a global body dedicated to developing voluntary, consensus-based standards that enhance quality, safety, and efficiency across industries. Initially focused on mechanical engineering—its first standard, ISO/R 1:1951, standardized industrial length measurements—ISO's scope expanded over decades to address diverse domains, from shipping container standardization in the 1960s to environmental management with ISO 14000 in the 1990s [24]. Today, with 169 member countries and over 24,000 standards, ISO facilitates international trade, reduces technical barriers, and supports regulatory compliance. Its evolution into technology sectors, including artificial intelligence (AI), reflects a response to emerging global challenges, such as cybersecurity and ethical AI governance [25]. ISO's collaborative, market-driven approach ensures standards reflect stakeholder needs, making it a cornerstone for aligning AI innovation with societal values, a role this paper leverages to assess ethical risk mitigation.

### B. ISO's Approach to Standard Development in Technology

ISO's approach to developing technology standards is characterized by collaboration, market relevance, and technical rigor, ensuring adaptability to complex fields like AI [5]. It convenes experts from industry, academia, government, and consumer groups worldwide, fostering diverse input through technical committees. For instance, proposals are market-driven, responding to technological trends and societal needs, as seen in the rapid development of

AI standards post-2018. The process involves rigorous drafting, public consultation, and iterative revisions, ensuring standards like ISO/IEC 24028 [26] on AI trustworthiness are technically sound and widely applicable [27]. Regular updates maintain relevance, with committees like JTC 1 adapting to AI's evolving risks [28]. This methodical approach underpins ISO's ability to guide ethical AI development, though its voluntary nature poses challenges in enforcement, a limitation this paper's framework seeks to address across regulatory contexts.

*C. ISO/IEC JTC 1/SC 42 Committee*

The ISO/IEC JTC 1/SC 42 [8] committee, established in 2017 under the Joint Technical Committee 1 (JTC 1) of ISO and the International Electrotechnical Commission (IEC), is the primary body for AI standardization (ISO/IEC, 2023). SC 42 addresses the AI ecosystem through specialized working groups: WG1 (foundational standards), WG2 (data and analytics), WG3 (trustworthiness), WG4 (use cases), and WG5 (computational approaches), alongside joint working groups like JWG1 (governance) with other subcommittees (ISO/IEC, 2023). It has produced key standards, such as ISO/IEC 23894 [12] for risk management and ISO/IEC 24368 [29] for ethical considerations, aligning AI with global norms [18]. SC 42 collaborates with external bodies, like IEEE and OECD, to ensure interoperability and ethical coherence [5]. As of 2025, its focus on trustworthiness and governance provides a robust baseline for this paper's Comparative Risk-Impact Assessment Framework, which evaluates these standards' effectiveness across diverse regulatory landscapes

IV. GLOBAL REGULATORY FRAMEWORKS

*A. Emerging AI Regulations*

The rapid proliferation of artificial intelligence (AI) has prompted diverse regulatory responses worldwide, each addressing distinct ethical risks like bias, transparency, and societal harm. This subsection surveys frameworks in ten key regions, setting the stage for the Comparative Risk-Impact Assessment Framework by highlighting varied approaches to ethical governance.

The United Kingdom pursues a "pro-innovation" approach, rooted in its 2023 AI Regulation White Paper [30], favoring a principles-based, sector-specific framework over broad legislation. Regulators like the Financial Conduct Authority (FCA) enforce ethical oversight, with the Consumer Duty (effective 2023) tackling fairness in AI-driven finance (FCA, 2023). The Information Commissioner's Office (ICO) ensures transparency via AI auditing [31]. The 2023 AI Safety Summit and the AI Safety Institute underscore global leadership, though the Labour government's 2024 consultation signals a shift toward mandatory high-risk AI rules.

The United States exhibits a fragmented approach, with federal deregulation under Trump's 2025 Removing Barriers to American Leadership in AI executive order prioritizing innovation over ethical safeguards [32]. State-level efforts contrast sharply: Colorado's AI Act targets bias in high-risk systems [13], Illinois' HB 3773 (2024) [33] addresses employment discrimination, and California's SB-942 (2026) mandates transparency for generative AI [13]. Tennessee's ELVIS Act (2024) [34] mitigates deepfake risks, highlighting varied ethical priorities [2].

Since launching its *New Generation AI Development Plan* [35] in 2017, China has aimed to balance AI innovation with regulatory control. It has introduced key frameworks such as the *Interim Measures for Generative AI Services* [36] and the *2023 Deep Synthesis Measures* to address algorithmic accountability and deepfake risks [10]. In 2024, the country released *Ethical Norms for AI* that emphasize social stability. This commitment was further reflected in its participation at the 2023 UK AI Safety Summit.

Canada has released Pan-Canadian AI Strategy [37], by which, through the Standards Council of Canada, the Government of Canada is supporting efforts to advance the development and adoption of standards related to AI. The Artificial Intelligence and Data Act (AIDA) [38] which is under review, targets accountability in high-impact AI, while the Personal Information Protection and Electronic Documents Act (PIPEDA) [39] ensures privacy.

Australia integrates ethics via the 2019 National AI Ethics Framework [40], emphasizing transparency across eight principles, supported by the 2021 AI Action Plan [41]. The ACCC and OAIC enforce consumer and privacy protections, with 2024 consultations proposing mandatory high-risk AI rules [19].

India advances ethical AI with the 2018 #AIforAll Strategy, targeting inclusivity in healthcare and education, guided by the 2021 Responsible AI Principles (NITI Aayog, 2021) [42]. The 2024 IndiaAI Mission [43] and forthcoming Digital India Act address bias in high-risk systems, bolstered by GPAI leadership [44].

Japan prioritizes societal harmony via the 2019 Social Principles of Human-Centric AI [45], focusing on safety and fairness, updated in the 2024 AI Strategy (METI, 2024). The 2024 AI Guidelines for Business [46] and Hiroshima AI Process [47] reflect a soft-law approach with global resonance [18].

Singapore leads with the 2019 Model AI Governance Framework [48], updated 2024 [49], tackling transparency in AI deployment, complemented by the 2024 Generative AI Evaluation Sandbox. The National AI Strategy 2.0 (2023) [50] aligns ethics with innovation.

The Ministry of Science and ICT (MSIT) of South Korea announced the passage of the Basic Act on Artificial Intelligence (AI Basic Act) [51] by the National Assembly on December 26, 2024, set to take effect in January 2026. This landmark legislation, making Korea the second country globally to enact such a law, aims to bolster national AI competitiveness and establish a trustworthy AI framework amid growing global competition, following models like the EU AI Act (2024). Developed over four years through bipartisan efforts, the Act outlines a national AI governance structure, including a triennial AI Master Plan, the National AI Committee chaired by the President, and an AI Safety Institute. It supports the AI industry through R&D, data training, AI clusters, and talent development, while addressing risks from high-risk and generative AI with transparency, safety, and operator accountability measures, positioning Korea to become a top global AI leader.

Brazil has made steady progress in shaping its AI governance landscape. Initially, Bill 21/2020, introduced in 2020 and approved by the House of Representatives in 2021, set forth a proposed legal framework for artificial intelligence, emphasizing transparency and accountability in high-risk systems. This framework evolved through legislative deliberations and was enacted as the Brazil AI Act in 2024 [52], incorporating guidelines for ethical deployment and oversight. Complementing this, Brazil's National AI Strategy 2021 [53], places a strong emphasis on inclusivity and fairness, aligning with global efforts to address algorithmic bias, as highlighted by Obermeyer et al. [21].

## B. The European Union's AI Act

The European Union Artificial Intelligence Act (AIA) [54], effective from August 2024, establishes a comprehensive, risk-based framework for ethical AI. This framework was informed by earlier analyses, such as Veale and Borgesius's 2021 study [14] of the draft AIA, which provided insights into its proposed scope, including its application to providers and deployers within and beyond the EU. The AIA bans unacceptable-risk systems like mass surveillance and imposes strict requirements on high-risk AI systems. For minimal-risk systems, it mandates transparency, a principle that resonates with the ethics-based auditing approach outlined by Mökander et al [11]. General-purpose AI systems face additional obligations if they pose systemic risks. Supported by regulatory sandboxes and national enforcement, the AIA integrates with existing laws to set a global benchmark for ethical AI governance. This paper uses the AIA as a reference to evaluate the alignment of ISO standards with these ethical principles.

## V. MAPPING OF ISO STANDARDS

### A. Data and Governance-Related Standards

Data quality and governance are essential for building trustworthy AI, a priority highlighted by the EU AI Act's demand for reliable datasets in high-risk systems. ISO/IEC 5259-1:2021 offers a framework for managing data quality in analytics and machine learning, focusing on accuracy, completeness, and consistency to prevent errors that might cause biased results. For example, it supports the EU's requirement for representative training data to lower the risk of discrimination. Additionally, ISO/IEC 38507:2022 [55] provides governance guidelines for organizations using AI, aligning with the EU's accountability rules by clarifying oversight roles and responsibilities. While these standards promote ethical data practices, they aren't mandatory, which poses challenges in regions with less strict regulations [11]. This connection helps the framework assess how data-related ethical risks vary across different global contexts.

### B. Ethics, Fairness, and Transparency-Related Standards

The EU AI Act, requires high-risk AI systems to be transparent and fair. These goals align with ISO standards that focus on ethical AI practices. For example, ISO/IEC 24368:2022 [29] sets out principles like fairness, transparency, and respect for human values, which support the EU's push for explainable AI and reducing bias. Likewise, ISO/IEC 24027:2021 offers ways to spot and fix unwanted bias in AI, addressing issues like those found in healthcare algorithms. While these standards fit well with EU rules—such as the need to label AI-generated content—their voluntary nature means they're not always enforced outside strict regulatory environments. This gap is a central concern for the Comparative Risk-Impact Assessment Framework, which examines how effectively these standards promote fairness and transparency worldwide..

### C. Risk Management-Related Standards

Risk management is central to the EU AI Act's classification of AI systems by risk level, requiring robust mitigation for high-risk applications. ISO/IEC 23894:2023 provides a comprehensive risk management framework for AI, guiding organizations in identifying, assessing, and treating risks like safety failures or privacy breaches (ISO/IEC 23894:2023). It aligns with the EU's human oversight and documentation mandates, providing a structured approach to compliance. Additionally, ISO/IEC 24028:2020 focuses on trustworthiness in AI through resilience and security, which is supported by research on Trustworthy AI [3]. This research emphasizes principles like beneficence and non-maleficence, essential for managing systemic risks in general-purpose AI as per the EU's provisions. While these standards are effective in structured regulatory environments like that of the EU, their flexibility may be insufficient in regions prioritizing different risks, such as social stability, a challenge addressed by our framework.

### D. Quality Management-Related Standards

Quality management ensures AI systems meet performance and ethical benchmarks, a priority under the EU AI Act's technical robustness requirements. ISO/IEC 25059:2023 establishes a quality model for AI-based systems, defining criteria like reliability and usability, which align with the EU's emphasis on system accuracy and safety. This standard supports compliance by offering evaluation metrics for high-risk AI, such as autonomous vehicles [7]. Paired with ISO/IEC 5338:2023, which provides lifecycle management for AI development, it ensures consistent quality from design to deployment. Yet, their emphasis on technical quality might overlook societal impacts like equity—an issue Mittelstadt flagged [19] in his critique of principle-based AI ethics. The Comparative Risk-Impact Assessment Framework explores this gap across global regulatory contexts.

## VI. THE COMPARATIVE RISK-IMPACT ASSESSMENT FRAMEWORK

### A. Framework Description

The Comparative Risk-Impact Assessment Framework is a novel tool designed to evaluate and enhance the effectiveness of ISO AI standards in mitigating ethical risks across diverse global regulatory contexts. It addresses the limitation of static standard mappings by systematically analyzing how well ISO standards align with region-specific ethical priorities, offering a prescriptive approach to strengthen trustworthiness in AI systems [11]. The framework comprises three components:

Risk Identification: Identifies key ethical risks—bias, transparency, accountability, privacy, and societal harm—drawn from ISO/IEC 24368 and global regulations (ISO/IEC, 2023a; [44].

Impact Scoring: Assesses each risk's severity (1-5) and likelihood (1-5) within a regulatory context, producing a composite impact score (severity × likelihood, max 25), reflecting local ethical and legal priorities [3].

Standards Alignment: Scores ISO standards' mitigation effectiveness (1-5) for each risk, based on their scope, specificity, and enforceability [18].

For example, bias in hiring AI might score high impact in a region with anti-discrimination laws but lower where unregulated, while ISO/IEC 24027's voluntary bias tools might score moderately due to lack of mandates. This structured approach, visualized in Table 1 (below), enables gap analysis and tailored enhancements, advancing ISO's role in ethical AI governance beyond current mappings [14]

TABLE I. FRAMEWORK COMPONENTS EXAMPLE

| *Risk* | *Severity* | *Likelihood* | *Impact Score* | *ISO Standard* | *Mitigation Score* |
|---|---|---|---|---|---|
| Bias | 4 | 3 | 12 | 24027 | 3 |
| Transparency | 3 | 4 | 12 | 24368 | 4 |

*B. Application to Case Studies*

The framework is applied to three regulatory contexts to demonstrate its utility: the EU (EU AI Act), US-Colorado (Colorado AI Act), and China (Generative AI Measures).

European Union (EU AI Act): The EU's risk-based approach prioritizes bias and transparency in high-risk AI (European Commission, 2024). Bias scores high (Severity 4, Likelihood 4, Impact 16) due to strict anti-discrimination rules, with ISO/IEC 24027 scoring 4 for its robust bias assessment tools, aligning well with EU mandates [21]. Transparency (Severity 3, Likelihood 5, Impact 15) reflects disclosure requirements, matched by ISO/IEC 24368's explainability focus (Score 4) [18]. Accountability scores moderately (Impact 12), with ISO/IEC 23894 (Score 3) lacking enforceable oversight mechanisms [11].

United States-Colorado (Colorado AI Act): Colorado targets algorithmic discrimination in hiring and healthcare [13]. Bias is critical (Severity 5, Likelihood 3, Impact 15) due to legal mandates, but ISO/IEC 24027 scores only 2, as its voluntary nature limits enforcement in a deregulated federal context [2]. Transparency (Severity 3, Likelihood 3, Impact 9) is less emphasized, with ISO/IEC 24368 (Score 3) offering guidance but no binding rules. Accountability (Impact 10) aligns poorly with ISO/IEC 38507 (Score 2), highlighting governance gaps.

China (Generative AI Measures): China prioritizes social stability and privacy in AI content [10]. Privacy scores high (Severity 4, Likelihood 4, Impact 16) due to state oversight, but ISO/IEC 24028 (Score 2) lacks specific privacy controls, missing China's focus. Transparency (Severity 2, Likelihood 3, Impact 6) is secondary, with ISO/IEC 24368 (Score 3) partially relevant. Societal harm (Impact 18) is critical, yet ISO/IEC 23894 (Score 3) offers generic risk management, not tailored to stability concerns.

*C. Findings and Gap Analysis*

The framework reveals significant gaps in ISO standards' adaptability. In the EU, high alignment exists for bias and transparency (Scores 4), but accountability lags due to voluntary adoption (Score 3) (European Commission, 2024). Colorado shows weaker alignment, with bias mitigation (Score 2) and accountability (Score 2) undermined by ISO's lack of enforceability, exacerbating risks in a fragmented U.S. landscape [13]. China highlights a privacy gap (Score 2) and insufficient focus on societal harm (Score 3), reflecting ISO's Western-centric design. Across cases, voluntary standards struggle where regulations demand binding rules [19], and regional ethical priorities (e.g., stability in China) are underrepresented, limiting ISO's global effectiveness.

*D. Recommendations for Enhancing ISO Standards*

To address these gaps, we recommend some enhancements to ISO standards. Building on the ethical framework for a good AI society proposed by Floridi et al. [4] which emphasizes principles like data governance and transparency, and the trustworthy AI concept outlined by Thiebes et al. [3], focusing on beneficence and justice, we propose the following seven enhancements to ISO standards. These recommendations aim to mitigate ethical risks, enhance trust, and ensure interoperability worldwide by establishing foundational technical standards and integrating them into existing ISO frameworks:

- Development of an AI-Augmented Cybersecurity Standard: Create a new ISO standard focused on securing AI systems against adversarial threats. This standard will define technical frameworks, including methodologies for robustness testing against adversarial attacks, secure model deployment protocols with end-to-end encryption, and AI-driven threat detection mechanisms, to ensure AI systems are resilient and trustworthy.

- Development of an AI Edge Computing Standard: Establish a new ISO standard for AI in edge computing and federated learning systems. This standard will provide compatibility frameworks, including interoperability protocols for edge AI models, privacy-preserving machine learning techniques for decentralized training, and guidelines for secure operation in distributed environments, addressing efficiency and privacy needs.

- Mandatory Risk Audit Standard: Develop a new ISO standard mandating periodic risk audits for high-impact AI, incorporating the technical frameworks from the AI-Augmented Cybersecurity Standard (Recommendation 1) for robustness testing and secure deployment, to enhance accountability beyond the voluntary framework of ISO/IEC 23894 (e.g., addressing needs in the EU and Colorado).

- Region-Specific Annexes: Add annexes to standards like ISO/IEC 24368 and 24028, tailoring guidance [sic] guidance to regional risks—e.g., privacy for China, bias for Colorado—improving adaptability, supported by the AI Edge Computing Standard's (Recommendation 2) privacy-preserving techniques for region-specific compliance.

- Strengthened Privacy Module: Expand ISO/IEC 24028 with a privacy-specific module, aligning with global data protection trends (e.g., China's needs), reinforced by the AI Edge Computing Standard's (Recommendation 2) federated learning protocols and the AI-Augmented Cybersecurity Standard's (Recommendation 1) secure model training guidelines to ensure robust privacy protections.

- Integration of AI-Augmented Cybersecurity Frameworks: Incorporate the AI-Augmented Cybersecurity Standard (Recommendation 1) into existing ISO frameworks, applying its standardized methodologies for robustness testing, secure deployment, and AI-driven threat detection to complement the risk audit standard and enhance overall system trustworthiness.

- Integration of AI Edge Computing Frameworks: Integrate the AI Edge Computing Standard (Recommendation 2) into ISO frameworks, utilizing its compatibility protocols and privacy-preserving techniques to support the privacy module and ensure secure, efficient AI operation in distributed environments.

These recommendations ensure ISO standards better mitigate ethical risks, enhancing trust and interoperability worldwide [3].